\documentclass[12pt,epsf]{article}
\usepackage{graphicx}
\usepackage{epsfig}
\begin{document}

\def\a{\alpha}
\def\b{\beta}
\def\c{\varepsilon}
\def\d{\delta}
\def\e{\epsilon}
\def\f{\phi}
\def\g{\gamma}
\def\h{\theta}
\def\k{\kappa}
\def\l{\lambda}
\def\m{\mu}
\def\n{\nu}
\def\p{\psi}
\def\q{\partial}
\def\r{\rho}
\def\s{\sigma}
\def\t{\tau}
\def\u{\upsilon}
\def\v{\varphi}
\def\w{\omega}
\def\x{\xi}
\def\y{\eta}
\def\z{\zeta}
\def\D{\Delta}
\def\G{\Gamma}
\def\H{\Theta}
\def\L{\Lambda}
\def\F{\Phi}
\def\P{\Psi}
\def\S{\Sigma}

\def\o{\over}
\def\beq{\begin{eqnarray}}
\def\eeq{\end{eqnarray}}
\newcommand{\gsim}{ \mathop{}_{\textstyle \sim}^{\textstyle >} }
\newcommand{\lsim}{ \mathop{}_{\textstyle \sim}^{\textstyle <} }
\newcommand{\vev}[1]{ \left\langle {#1} \right\rangle }
\newcommand{\bra}[1]{ \langle {#1} | }
\newcommand{\ket}[1]{ | {#1} \rangle }
\newcommand{\EV}{ {\rm eV} }
\newcommand{\KEV}{ {\rm keV} }
\newcommand{\MEV}{ {\rm MeV} }
\newcommand{\GEV}{ {\rm GeV} }
\newcommand{\TEV}{ {\rm TeV} }
\def\diag{\mathop{\rm diag}\nolimits}
\def\Spin{\mathop{\rm Spin}}
\def\SO{\mathop{\rm SO}}
\def\O{\mathop{\rm O}}
\def\SU{\mathop{\rm SU}}
\def\U{\mathop{\rm U}}
\def\Sp{\mathop{\rm Sp}}
\def\SL{\mathop{\rm SL}}
\def\tr{\mathop{\rm tr}}

\def\IJMP{Int.~J.~Mod.~Phys. }
\def\MPL{Mod.~Phys.~Lett. }
\def\NP{Nucl.~Phys. }
\def\PL{Phys.~Lett. }
\def\PR{Phys.~Rev. }
\def\PRL{Phys.~Rev.~Lett. }
\def\PTP{Prog.~Theor.~Phys. }
\def\ZP{Z.~Phys. }

\newcommand{\bear}{\begin{array}}  
\newcommand {\eear}{\end{array}}
\newcommand{\la}{\left\langle}  
\newcommand{\ra}{\right\rangle}
\newcommand{\non}{\nonumber}  
\newcommand{\ds}{\displaystyle}
\newcommand{\red}{\textcolor{red}}
\newcommand{\mwino}{m_{\widetilde{W}^0}}
\def\ubl{U(1)$_{\rm B-L}$}
\def\REF#1{(\ref{#1})}
\def\lrf#1#2{ \left(\frac{#1}{#2}\right)}
\def\lrfp#1#2#3{ \left(\frac{#1}{#2} \right)^{#3}}
\def\OG#1{ {\cal O}(#1){\rm\,GeV}}


\baselineskip 0.7cm

\begin{titlepage}

\begin{flushright}
IPMU 09-0054
\end{flushright}

\vskip 1.35cm
\begin{center}
{\large \bf
R-violating Decay of Wino Dark Matter and electron/positron Excesses in 
the PAMELA/Fermi Experiments
}
\vskip 1.2cm
Satoshi Shirai$^{1,2}$, Fuminobu Takahashi$^2$ and  T. T. Yanagida$^{2,1}$
\vskip 0.4cm

{\it $^1$  Department of Physics, University of Tokyo,\\
     Tokyo 113-0033, Japan\\
$^2$ Institute for the Physics and Mathematics of the Universe, 
University of Tokyo,\\ Chiba 277-8568, Japan}

\vskip 1.5cm

\abstract{ We show that R-parity violating decay of Wino dark matter
  of mass $\sim 3$ TeV can naturally account for the flux and spectral
  shape of the cosmic-ray electrons and positrons observed by the
  PAMELA and Fermi satellites. To provide a theoretical basis for the
  scenario, we also present a model that trilinear R-parity breaking
  appears with a coefficient suppressed by powers of the gravitino
  mass, which naturally leads to the Wino lifetime of $O(10^{26})$
  sec.  }
\end{center}
\end{titlepage}

\setcounter{page}{2}

\section{Introduction}
\label{sec:1}

The lightest supersymmetry (SUSY) particle called as LSP in the SUSY
standard model (SSM) is known as a good candidate of dark matter (DM)
in the Universe if its mass is $O(100)$\,GeV $-$ $O(1)$\,TeV.  The
stability of the LSP can be guaranteed by assuming an exact R
parity. However, the R parity may not necessarily be an exact
symmetry. In fact, the LSP can be still DM as long as R-parity
breaking terms are sufficiently small and the lifetime of the LSP is
much longer than the present age of the Universe. If this is the case,
the decay of DM may give rise to some excesses in cosmic rays.

Much attention was recently attracted to the anomalies in cosmic-ray
electron/positron fluxes observed by PAMELA~\cite{Adriani:2008zr} and
ATIC~\cite{:2008zz}, and decaying LSP scenarios were extensively
discussed in this context~\cite{Takayama:2000uz,gravitinoLSP,
  Yin:2008bs}.\footnote{
See also Ref.~\cite{decayingDM} for other decaying DM models.
}
Most of the proposals assume bilinear R-parity breaking terms such as
$LH_u$, since they are the lowest dimensional R-parity breaking
operators which most likely dominate R-parity breaking effects at low
energies. However, the bilinear R-parity breaking terms induce DM
decays into quarks and hence produce too many antiprotons in cosmic
rays. Therefore, it is much safer to consider the next lowest
dimensional operators such as $\bar{e}
LL$~\cite{Yin:2008bs,Hamaguchi:2008ta,Gogoladze:2009kv,Ishiwata:2009vx}.
However, it seems very unnatural to consider the second lowest
dimensional operators, suppressing the lowest dimensional bilinear
operators. In addition, even if the trilinear term dominates over the
bilinear one, the magnitude of the R-parity breaking should
be extremely small, especially in the case of neutralino LSP.  No
rigorous explanation for the smallness was known, and the size of the
R-parity violation was treated as a free parameter.

Very recently, the Fermi collaboration has released data on the
electron/positron fluxes from $20$\,GeV up to $1$\,TeV~\cite{Collaboration:2009zk};
the spectrum falls as $E^{-3.0}$ without prominent spectral features,
and it is in agreement with the H.E.S.S. data at $E \sim
1\,$TeV \cite{Collaboration:2008aaa,Aharonian:2009ah}.\footnote{ The Fermi result is not consistent with the excess
  around $E \simeq 600$\,GeV reported by ATIC.  In this letter we
  adopt the Fermi result and do not consider the ATIC data.  }  The
index of the observed electron/positron spectrum is close to the high
end of theoretically expected value. Moreover, if we combine the Fermi
data for $E \lsim 1$\,TeV and the H.E.S.S. data for $E \gsim 1$\,TeV, it
looks that the spectrum becomes softer at energies above $1\,$TeV.
From this observation, we regard the relatively hard (and almost
featureless) electron/positron spectrum below $1$\,TeV reported by
Fermi as an excess with respect to the background.  If the
electron/positron spectrum observed by Fermi (as well as the positron
fraction observed by PAMELA) is to be explained by the DM decay , the
mass of DM must be about a few TeV. The suggested mass scale is
intriguingly close to the mass of a Wino LSP required for the thermal
relic to explain the observed DM abundance.

In this letter we present a model that trilinear R-parity breaking
appears with a coefficient suppressed by powers of the gravitino mass.
Interestingly, the size of the R-parity breaking is naturally small,
and moreover, it leads to the required size of the R-parity breaking
suggested by observation, in the case of the Wino LSP of a mass $\sim
3$\,TeV.  Based on this model, we show that the Fermi as well as
PAMELA data can be simultaneously explained by the Wino LSP decaying
through the trilinear term $\bar{e} LL$.

\section{A model of R-parity breaking}
\label{sec:2}

It is known that the superpotential possesses a constant term $C_0$ to
cancel the positive energy density induced by SUSY breaking.  The
constant term is equal to the gravitino mass, $C_0 = m_{3/2}$, in the
Planck unit in which the Planck mass $M_P \simeq 2.4\times 10^{18}$
GeV is set to be unity.  The constant term breaks the continuous
$U(1)_R$ symmetry down to a discrete $Z_2$ symmetry, which is nothing
but the R parity. A crucial observation is that, if the R symmetry at
high-energies is not the continuous $U(1)_R$ but a discrete
$Z_{2k+1}~(k={\rm integer)}$, the constant term $C_0$ 
results in the R-parity breaking. In this letter we take $k=2$, since, as we will see
later, it leads to the required magnitude of the R-parity breaking
operators to account for the anomalous excess in cosmic-ray electrons
and positrons.

Let us take the R charges for all quark and lepton chiral multiplets
to be $1$ and those for the Higgs chiral multiplets $H_u$ and $H_d$ to
be $0$. (See Table 1.) Then, the following lepton-number violating
trilinear term is allowed by the $Z_{5R}$ symmetry:
\begin{equation}
W\; =\; \kappa_{ijk} (C_0)^2 \bar{e}_i L_j L_k,
\label{eq:trilinear}
\end{equation}
where $\kappa_{ijk}$ is a numerical coefficient, $i,j,k=1,2,3$ denote
the generation, and summation over the SU(2) gauge indices is
understood. The coupling $\kappa_{ijk}$ must be antisymmetric under
the exchange of the last two indices ($j \leftrightarrow k$) due to
the SU(2) gauge invariance.  Note that the R charge of $C_0 (=
m_{3/2}) $ is $2$. We may presume that the coefficient $\kappa_{ijk}$
takes a larger value for the third (and second) generation. Therefore,
in the following we focus on $\kappa_{i 2 3}$ with $i=1,2,3$, which is
assumed to be unsuppressed with the other terms of different
combination of flavors being suppressed.

\begin{figure}[t]
\begin{center}
\epsfig{file=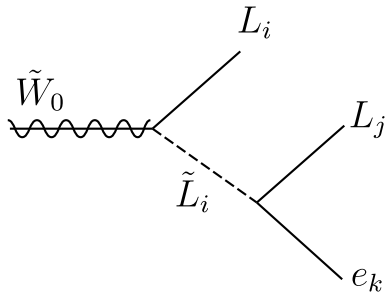 ,scale=.80,clip}\\
\end{center}
\caption{One of the diagrams representing the Wino decay  through the trilinear R-parity
breaking (\ref{eq:trilinear}).}
\label{fig:diagram}
\end{figure}

We assume that a Wino LSP accounts for the DM in the Universe.  In
order to explain the observed DM abundance, the Wino mass must be in
the range between $2.7 - 3$\,TeV~\cite{Hisano:2006nn}.  The Wino LSP
is naturally realized in the anomaly-mediated SUSY
breaking~\cite{AMSB}.  Then, using the relation between the Wino mass
and the gravitino mass in the anomaly mediation, we find that the
gravitino mass should be about $10^3$\,TeV.  In the presence of the
R-parity breaking (\ref{eq:trilinear}), the Wino LSP is no longer
stable, and decays into neutrinos and charged leptons through the
exchange of a virtual slepton.  One of the decay diagrams is shown in
Fig.~\ref{fig:diagram}.  The decay rate of the Wino LSP through the
interaction (\ref{eq:trilinear}) with $(j,k)=(2,3)$ is given
by~\cite{Baltz:1997gd}
\beq
\Gamma(\widetilde{W}^0 \rightarrow \tau^{\pm}\nu_{\mu} e_i^{\mp},\nu_{\tau}\mu^{\pm}e_i^{\mp})  
\! \sim \!\left(10^{27} {\rm sec}\right)^{-1} |\kappa_{i23}|^2
\lrfp{m_{3/2}}{10^3~{\rm TeV}}{4} \lrfp{\mwino}{3~{\rm TeV}}{5} \lrfp{m_{\tilde{\ell}}}{5{\rm \, TeV}}{-4},
\label{lifetime}
\eeq
where $m_{\widetilde{W}^0}$ denotes the Wino mass, and we have assumed
the common slepton mass, $m_{\tilde{\ell}}$, for simplicity. As we will
see in the next section, the lifetime (\ref{lifetime}) is close to
what is needed to explain the cosmic-ray observation.

Several comments are in order. First, there could be other trilinear
terms including quark multiplets, but they may have only negligible
effects on the decay processes if squarks are substantially heavier
than sleptons. Second, the bilinear term is also allowed by the
$Z_{5R}$ symmetry and takes the following form:
\beq
W\;=\; (C_0)^3LH_u.
\label{bilinear}
\eeq
Since the bilinear term is suppressed by an additional factor $C_0$
compared to the trilinear term, the latter becomes much more effective
than what is naively expected based on the dimensional grounds.  It is
not trivial, though, if the trilinear term dominates over the bilinear
term as the decay processes of the LSP.  We will come back to this
issue in Sec.~\ref{sec:concl}, and for the moment, we simply assume
that the bilinear term does not have sizable effects on the decay.

\begin{table}[t]
\label{rcharge}
\begin{center}
\begin{tabular}{|c||c|c|c|c|c|c|c|c|c|}
  \hline
  &$Q$ & $\bar{u}$ & $L$ &$\bar{e}$ &$H_u$ &$H_d$&$N$ &$M$&$C_0$\\\hline
  R  & 1         &     1        &       1       &           1           &        0              &       0  &1 &0&2\\  \hline            
  $Z_3$& 1         &     1        &       1       &           1           &        1             &       1 &1 &1&0\\  \hline            
\end{tabular}
\caption{The  assignment of R-charge and $Z_3$. }
\end{center}
\end{table}

\section{Cosmic-ray signal from LSP decay}
\label{sec:CR}
\begin{figure}[t]
\begin{tabular}{cc}
\begin{minipage}{0.5\hsize}
\begin{center}
\epsfig{file=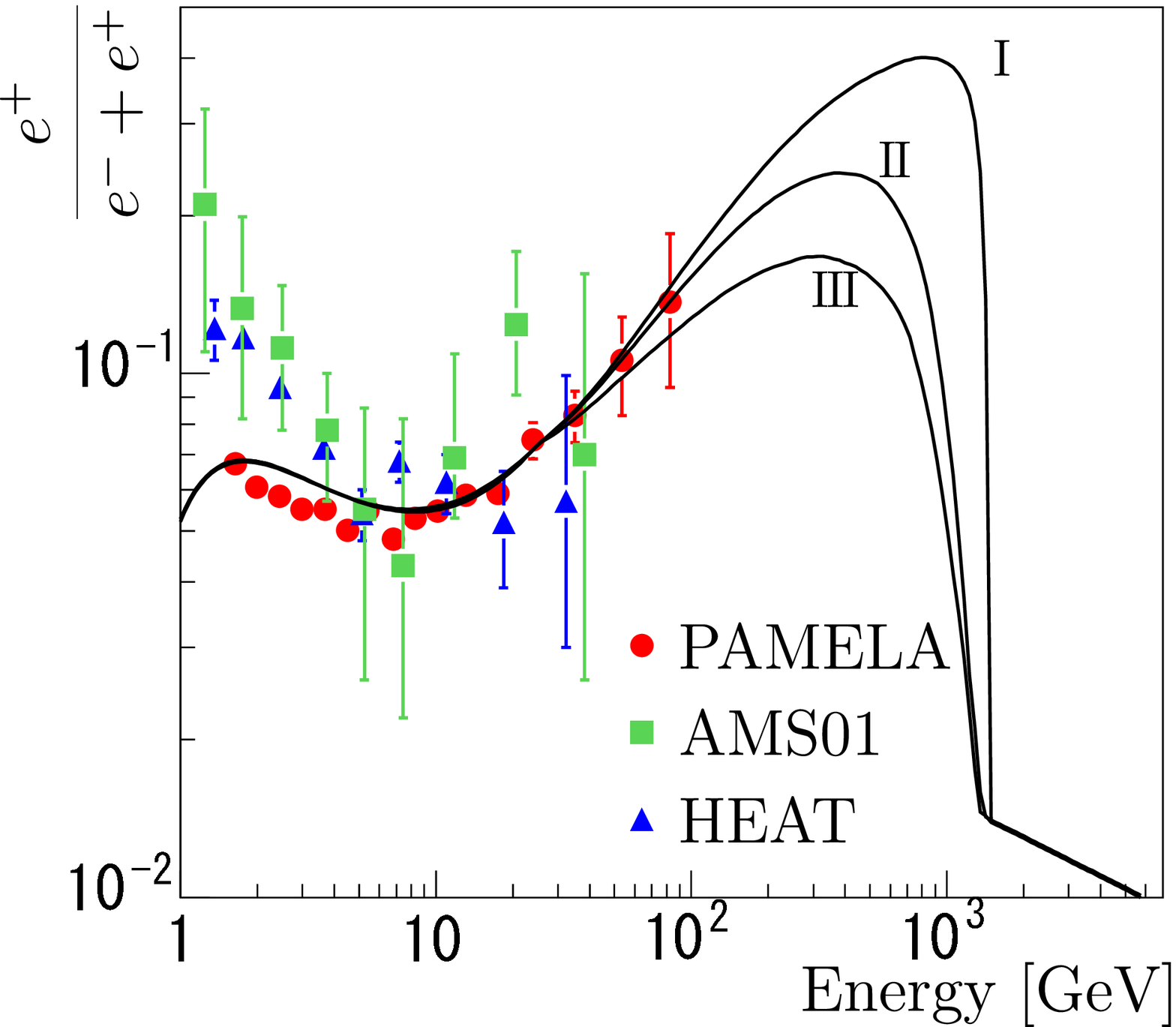 ,scale=.42,clip}\\
(a)
\end{center}
\end{minipage}
\begin{minipage}{0.5\hsize}
\begin{center}
\epsfig{file=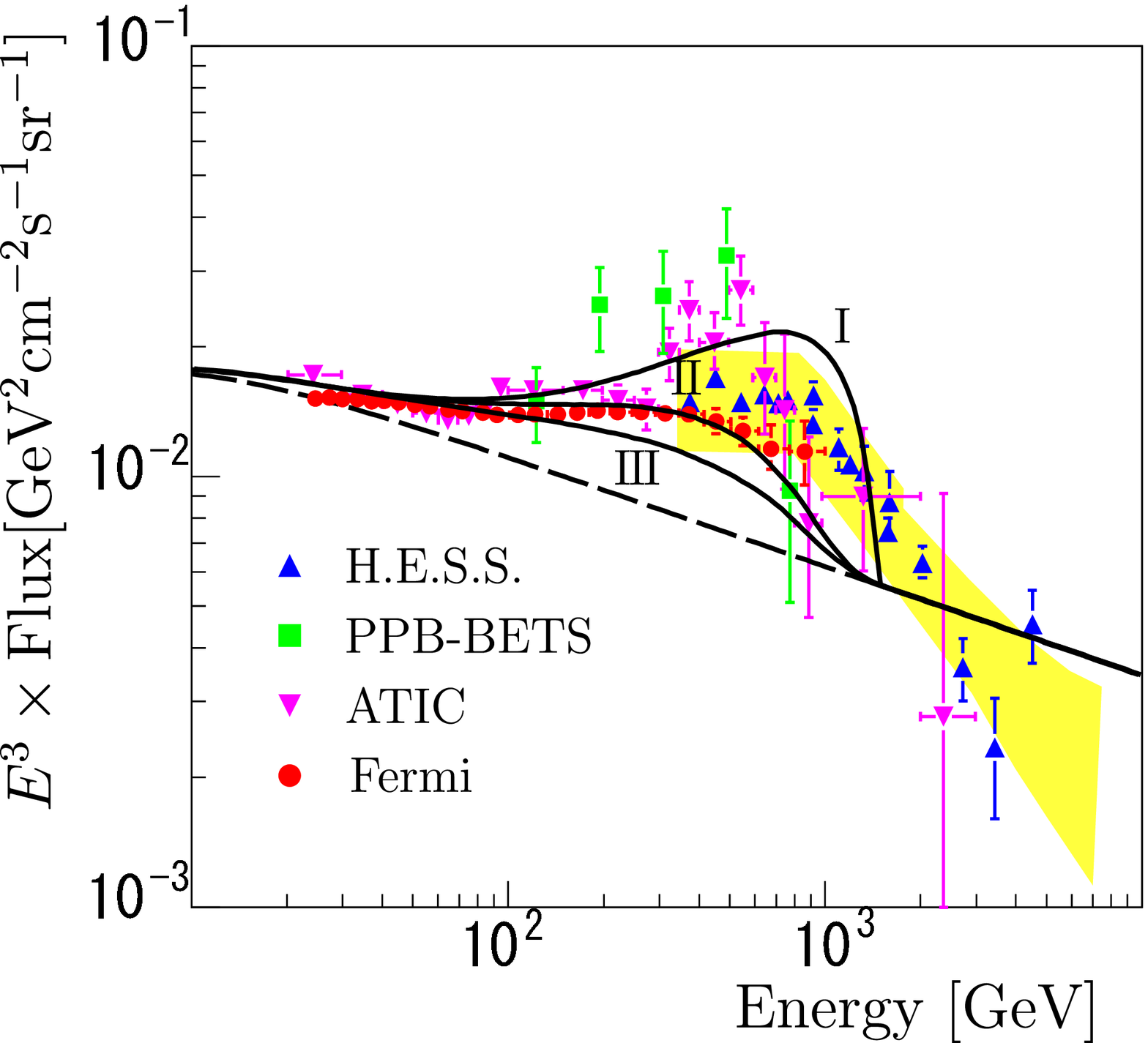 ,scale=.42,clip}\\
(b)
\end{center}
\end{minipage}\\
\end{tabular}
\caption{Cosmic ray signals in the present model.
(a): positron fraction with experimental data \cite{Adriani:2008zr,Aguilar:2007yf,Barwick:1997ig}.
(b): positron and electron fluxes with experimental data~\cite{Collaboration:2009zk,Collaboration:2008aaa,Aharonian:2009ah,:2008zz,Torii:2008xu}.
The yellow zone shows a systematic error and the dashed line shows the background flux.
I, II and III represent the cases that $\bar{e}_{1} L_2L_3$, $\bar{e}_{2} L_2L_3$ and $\bar{e}_{3} L_2L_3$ dominate the R-breaking, respectively.
}
\label{fig:signal}
\end{figure}

Let us discuss the cosmic ray signals from the LSP decay.  Its decay
pattern depends on R-breaking structure and the SSM mass spectrum.
For a demonstration, we consider the case that the $\bar{e}_{i}
L_2L_3~(i=1,2,3)$ term dominates the R-breaking and assume that ${\rm
  BF}({\rm DM} \to \tau^{\pm}\nu_{\mu} e_i^{\mp})={\rm BF}( {\rm DM}
\to\nu_{\tau}\mu^{\pm}e_i^{\mp})=0.5$.  We have used the constant
matrix element in the three-body phase space, for simplicity.  The
electron and positron energy spectrum is estimated with the program
PYTHIA~\cite{Sjostrand:2006za}.  For the propagation of the cosmic ray
in the Galaxy, we adopt the same set-up in Ref.~\cite{Shirai:2009kh},
based on Refs.~\cite{Hisano:2005ec,Ibarra:2008qg}. As for the electron
and positron background, we have used the estimation given in
Refs.~\cite{Moskalenko:1997gh,Baltz:1998xv}, with a normalization
factor $k_{\rm bg} = 0.68$.  In Fig.~\ref{fig:signal}, we show the
positron fraction and the electron and positron total flux.  We set
that $m_{\rm DM}= 3$ TeV and the lifetime is $9\times 10^{25}$ sec. As
can be seen from Fig.~\ref{fig:signal}, the cosmic-ray signal in the
present model can nicely fit the PAMELA data for $i=1,2,3$. On the
other hand, the prediction in case of $i=1$ fails to explain the Fermi
data due to the presence of hard electrons produced by the LSP decay,
leading to the bump at $1$\,TeV. The prediction in case of $i=2$ gives
a very good fit to the Fermi data. Considering uncertainties in the
background estimation as well as the SSM mass spectrum, however, the
case of $i=3$ may be able to give an equally good fit.

\section{Discussion and conclusions}
\label{sec:concl}
Let us discuss several issues in the model presented in
Sec.~\ref{sec:2} and possible solutions. First of all, the bilinear
term like (\ref{bilinear}) could induce additional decay processes
into lepton and Higgs, which may result in too many antiprotons. Note
that the effective mixing angle induced by the bilinear term is given
by the ratio of $C_0^3$ to the Higgs mass or slepton mass. Therefore,
if $m_{3/2}$ is the same orders of magnitude of the Higgs mass or the
slepton mass, the bilinear term will be as important as the trilinear
term.

To make the matter worse, the bilinear term will be enhanced in the
presence of right-handed neutrinos $N$.  This is because, if we assign
the R-charge $1$ to $N$ as in Table 1, the following interactions are
allowed by the $Z_{5R}$ symmetry.
\beq
W = C_0^3 N +\frac{1}{2} M NN + y_\nu N L H_u,
\label{N}
\eeq
where $M$ denotes the Majorana right-handed neutrino mass, $y_\nu$ is
the neutrino Yukawa coupling, and we have suppressed the flavor
indices.  Since the first term is generically present, the
right-handed neutrino will develop a non-vanishing expectation value
of $O(C_0^3/M)$.  Then, the neutrino Yukawa coupling induces the
bilinear term $y \la N \ra L H_u \sim (C_0^3/M) LH_u$, which is
enhanced by $1/M$ compared to (\ref{bilinear}). Thus, the presence of
the right-handed neutrino of mass lighter than the Planck mass leads
to the enhancement of the bilinear term.

Those problems can be easily solved by introducing a $Z_3$ symmetry,
under which the SSM particles and $M$ are charged by an unit
charge. (See Table 1.)  Namely, the $Z_3$ symmetry is broken by $M \ll
1$. Then the bilinear term (\ref{bilinear}) is suppressed by $M$,
while the first term in (\ref{N}) by $M^2$. Therefore, the trilinear
term (\ref{eq:trilinear}) will be the dominant source of the LSP
decay.

In this letter we have presented the R-parity breaking model in which
the trilinear term becomes important and appears with a coefficient
proportional to powers of the gravitino mass. We have also shown that
the Wino LSP decaying through the lepton-number violating trilinear
term (\ref{eq:trilinear}) can account for the PAMELA and Fermi data.
 
There are several non-trivial coincidence in our scenario.  First of
all, the change in the power index of the electron/positron spectrum
suggests the DM of mass a few TeV. This is intriguingly close to the
mass of the thermal relic Wino DM. Assuming the anomaly mediation, the
gravitino mass is determined to be about $10^3$\,TeV. In our model on
the R-parity violation, the lifetime of the Wino LSP is determined by
some combination of the Wino mass and the gravitino mass.
Substituting the above values for the Wino and gravitino masses, we
have obtained the lifetime which is surprisingly close to what is
needed to account for the cosmic-ray anomalies.  Further study of the
prediction of other cosmic rays such as gamma-rays, antiprotons and neutrinos\footnote{
Our model is consistent with the current observational bound on the neutrino production
from decaying DM~\cite{Hisano:2008ah}.
}, as
well as future observational data, will enable us to tell whether
those are just coincidence or may reflect from the characteristics of
DM and the underlying physics beyond the SM.

\begin{figure}[t]
\begin{center}
\epsfig{file=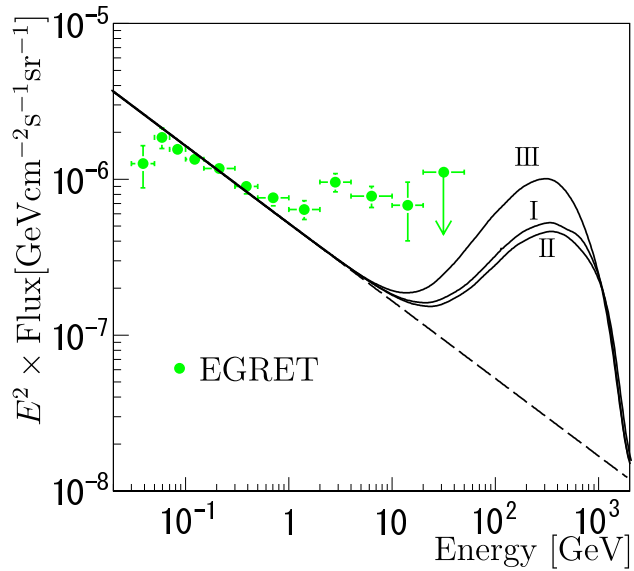 ,scale=.43,clip}\\
\end{center}
\caption{Predicted signals of diffuse gamma-ray flux shown together 
with the EGRET data~\cite{Sreekumar:1997un,Strong:2004ry}.}
\label{fig:diffuse-g}
\end{figure}

We can see from Fig.~\ref{fig:signal} that the fit to the Fermi data becomes better
if the decay products include the charged leptons in the second and
third generations, namely,  muons and taus. In particular, if the decay product is dominated by
the muon (as in the case II of Fig.~\ref{fig:signal}), the fit looks pretty
good.\footnote{ See Ref.~\cite{IMSY} for a model in which the
muons are mainly produced by the DM decay.  } It is actually
possible to give a equally nice fit to the Fermi data, if we properly combine
the contributions from the first and third generations. This is indeed the case if we consider
unsuppressed $\kappa_{313}$. Such flavor dependence
may be probed by studying the diffuse gamma-ray in detail. For reference
we show in Fig.~\ref{fig:diffuse-g} the predicted diffuse gamma-ray signals for the cases I, II and III.
Including more taus in the final states generically lead to larger signal in the diffuse
gamma-ray. We may be able to untangle the flavor dependence by making use of
the different predictions.

\section*{Acknowledgement}
We thank Junji Hisano and Hitoshi Murayama for discussion.  This work was supported by
World Premier International Center Initiative (WPI Program), MEXT,
Japan.  The work of SS is supported in part by JSPS Research
Fellowships for Young Scientists.

\end{document}